# Charge and orbital ordering in Fe and Mn perovskite oxides far from half- doping by resonant x- ray scattering

**J Herrero-Martín[1], J García[2], G Subías[2], J Blasco[2], C Mazzoli[1] and M C Sánchez[2]**

[1] European Synchrotron Radiation Facility, BP 220, 38043 Grenoble Cedex 9, France
[2] Departamento de Física de la Materia Condensada, Instituto de Ciencia de Materiales de Aragón, CSIC-Universidad de Zaragoza, Pedro Cerbuna 12, 50009 Zaragoza, Spain

herrerom@esrf.fr

**Abstract**. The emergence of superlattice periodicities at metal to insulator transitions in hole doped perovskite oxides responds to a rearrangement of the local atomic structure, and electron and spin density distribution. Originally, the ionic model based on a checkerboard- type atomic distribution served to describe the low temperature charge and orbital ordered (COO) phases arising in half- doped manganites. In the last years, the exploitation of resonant x- ray scattering (RXS) capabilities has shown the need to revisit these concepts and improve the picture. Yet, we have realised that COO is a more common phenomenon than expected that can be observed in a wide range of doping levels. Here we compare the experimental data recently collected by RXS on $La_{0.4}Sr_{1.6}MnO_4$ (x=0.60) and $La(Pr)_{1/3}Sr_{2/3}FeO_3$ (x=0.67). The first shows a COO phase similar to that found in the x=0.5 sample but angular peak positions vs. T denotes the incommensurability of superlattice reflections. Meanwhile, the analysis of the commensurate CO phase in the studied ferrite underlines the role of the structural changes also involving La(Sr) and O atoms.

## 1. Introduction
The metal- insulator (MI) phase transition in transition metal (TM) oxides continues to draw the attention of the scientific community in condensed matter physics. Within the assumed linkage of this process to lattice structure, charge distribution and spin and orbital magnetic moments, the precise role of the any of the corresponding order parameters has not been fully detailed yet. Among these materials, Mn- based compounds are one of the most studied families due to the interest on the colossal magnetoresistive properties they can exhibit. An ionic model was proposed by Goodenough [1] to explain the different magnetic structures present in $La_{1-x}A_xMnO_3$ (A: alkali- earth) as a function of hole doping (x), i.e. the relation of $Mn^{3+}$: $Mn^{4+}$ ions. For the particular case of x=0.5, this ratio is 1:1 and would facilitate their periodic array in the crystal lattice in the low temperature charge localised phase thus leading to the so- called charge ordering (CO) and orbital ordering (OO) phenomena. Since this figure reproduced reasonably well many of the macroscopic properties of these materials it did not significantly changed so far. Nevertheless, the development of new experimental techniques such as the resonant x- ray scattering (RXS) currently allows us to characterise the local geometry and electronic structure of TM atoms in charge localised phases, combining the atomic selectivity that provides x- ray absorption and crystal sites contrast characteristic of diffraction [2]. The pioneer works on half- doped perovskites such as $La_{0.5}Sr_{1.5}MnO_4$ and $Nd_{0.5}Sr_{0.5}MnO_3$ [3,4] confirmed the existence of two well differentiated sites occupied by two Mn ions (approximated as $Mn^{3+}$ and $Mn^{4+}$) in the CO phase (below ~200 to 250 K). In agreement with previous results by x- ray diffraction [5], they showed that these sites order in the crystal lattice following zig-zag chains (as a checkerboard) propagating in the ab plane and leading to the appearance of a superlattice modulation originated in the charge segregation. Associated to it and with a doubled periodicity, single occupied $e_g$ orbitals of $Mn^{3+}$ ions would order following a CE- type scheme. This model has been generalised for half- doped $R_{1-x}A_xMnO_3$ (R: lanthanide), layered (or ~2D structure) $R_{1-x}A_{1+x}MnO_4$ compounds, intermediate $R_{2-2x}A_{1+2x}MnO_7$ series and similarly to other Co or Ni based oxides [6,7]. Nevertheless, in the last years an increasing number of both theoretical and experimental works have cast doubts on the validity of the ionic approximation. There have been detected two main problems. First, the charge segregation between dissimilar TM sites has been found to be much smaller than one electron at the time that the covalence with nearest O atoms has been remarked [8-10]. And secondly, CO phases have been found in doped manganites out of half- doping condition where the number of crystallographic sites do not match with the calculated expected ratio of +3/+4 ionic states [11]. It is convenient to remark here that the term CO is commonly used in the broadest sense, i.e. it comprehends the cases of a charge segregation between sites, no matter its magnitude. In spite of these results, CO systems based on different TMs continue to be generally described in terms of integer ionic states ($TM^{n+}/TM^{m+}$), those where m-n=1 being the most widely studied. In this sense, focusing on systems showing CO phases with a ratio of sites ≠1 and/or m-n≠1 appears to be a direct way towards a better knowledge of the physics of electron localisation.

Electron and x- ray diffraction results on $La_{1-x}Ca_xMnO_3$ and $La_{1-x}Sr_{1+x}MnO_4$ have shown that the CO and OO modulations become incommensurate with the crystal lattice for x>0.5 [12,13] except for x~0.67 or 0.75, probably due to the fact that it can be expressed as a simple rational number (2/3, 3/4) and facilitates the array of two Mn sites with dissimilar oxidation state in a 1:2 or 1:3 relation. For 0.5<x<0.67, the modulation vector depends approximately linearly on x, which has been interpreted in terms of a charge density wave (CDW) of Mn $e_g$ electrons. Indeed, electron diffraction results have discarded the existence of charge stacking faults (i.e. a bimodal distribution of Mn sites) and indicated the presence of a superlattice with a uniform periodicity [14].

Charge ordering has also been subject of deep investigations in Fe oxides, where we can find a similitude with manganites. First studies by Mössbauer and neutron diffraction [15,16] showed that the T driven charge localised phases in $La_{1-x}A_xFeO_3$ (A: Ca, Sr) occur in a wide doping range (x<0.7). Further results obtained by transmission electron microscopy demonstrated that the superlattice propagation vector **q** direction is found to be locked to (1/2 0 0) for x<0.5 and parallel to the [111] cubic perovskite direction for x>0.5 [17]. For x=2/3, it becomes commensurate with a (1/3 1/3 1/3) periodicity and it was postulated to be originated in a …AABAAB… sequence of two differentiated Fe ions, identified as $A=Fe^{3+}$ and $B=Fe^{5+}$. Yet, the same propagation vector would be obtained if we consider $B'=Fe^{4+}$, leading to a …AB'B'AB'B'… sequence. As we show further through this paper, the charge localisation is here strongly coupled to the stabilisation of lattice distortions affecting not only Fe atoms but also R and O.

In this work we summarise resonant x- ray scattering data collected on the over- doped layered manganite $La_{0.4}Sr_{1.6}MnO_4$ (x=0.6) at the Mn K edge and $La(Pr)_{1/3}Sr_{2/3}FeO_3$ (x=2/3) at the Fe K edge. We analyse the CO and OO phases with respect to the half- doped and under- doped systems. In the first case, superlattice reflections are found for the wave vectors $q_{v,h\pm}$~(h±ε, h±ε, 0) and $q_{w,h\pm}$~(h±2ε, h±2ε, 0), where 2ε=1-x, h is integer and considering the tetragonal *I4/mmm* space group. In particular, we present data for h=1, 2. For ferrites, using the primitive cubic perovskite structure notation (though real structure is discussed later) $q'_{CO,h'}$=(h'/3 h'/3 h'/3)$_p$, equivalent to $q'_{CO,h'}$=(0 0 2h')$_{hex}$ in hexagonal coordinates (*R-3c* space group). Though expected to be equivalent, different spectra are recorded for H=2, 4, 5. A structural model here presented explains these divergences.

## 2. Experimental details
Single crystals of $La_{0.4}Sr_{1.6}MnO_4$ (LSMO-214) and $R_{1/3}Sr_{2/3}FeO_3$ (R: La, Pr) (L(P)SFO) were grown by the floating zone method as described elsewhere [18,19]. They were characterised by x- ray powder diffraction and electrical and magnetic measurements to control their quality. For RXS experiments, the Mn crystal was carefully cut along a surface normal to the [110] direction. Meanwhile, the ferrites were cut perpendicular to the $[111]_p$. Data were recorded at the ID20 beamline in the ESRF in Grenoble, France. Incident x- rays were monochromatised to energies close to the Mn and Fe K absorption edges by means of a Si (1 1 1) crystal. Samples were mounted on a four (+1) circles diffractometer working in vertical scattering configuration (i.e. incident σ polarization) and equipped with a He close- cycle refrigerator for low T measurements. The additional circle in the diffractometer serves for φ azimuthal angle rotation around the scattering vector **Q**. The origin corresponds to crystallographic [001] and $[-110]_p$ parallel to the incident beam direction for LSMO-214 and LSFO, respectively. Polarisation of the scattered radiation was analysed (σ' or π') by means of a Cu (2 2 0) (Mn K) and MgO (2 2 0) (Fe K) single crystals at ~45 deg., radiation collection being performed by an avalanche photodiode. All spectra have been corrected for absorption.

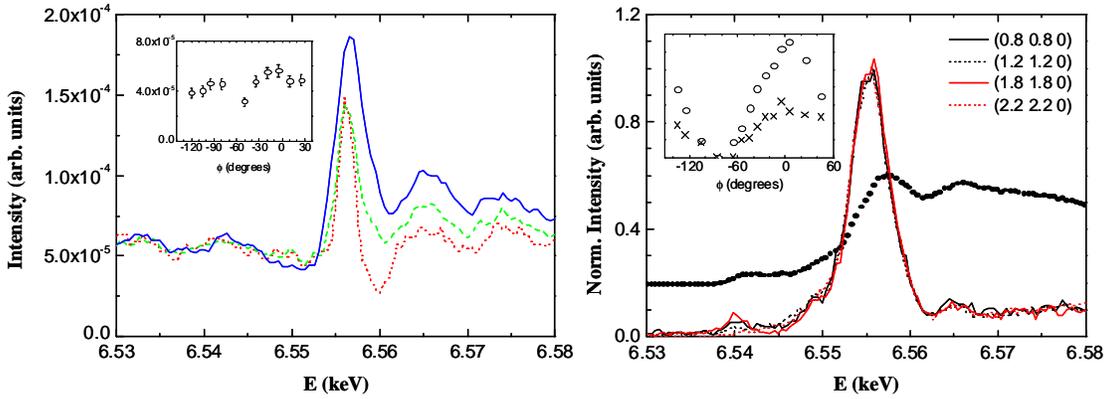

Figure 1. Left panel: $(1.6\ 1.6\ 0)_{\sigma\sigma'}$ at φ=-5° (solid line), -50° (dashed) and -80° (dotted). The inset shows the intensity at the maximum of the resonance as a function of φ. Right panel: $\mathbf{q}_{v,1(2)\pm}$ reflections in σ-π' channel. Fluorescence (circles) is also shown. Inset: azimuthal evolution at the pre- peak (crosses) and main resonance (circles). Minima are zero. All spectra have been recorded at 80 K.

## 3. Results and discussion

Energy dependent spectra for the $\mathbf{q}_{w,2-}$~(1.6 1.6 0) reflection in the σ-σ' channel and the four possible $\mathbf{q}_{vi\pm}$ (i=1,2) in the rotated σ-π' channel in LSMO-214 (ε=0.2) at the Mn K edge are shown in Figure 1. $I_{\sigma-\sigma'}$ was absent in the latter reflections while a very weak σ-π' contribution (near the detection limit) cannot be discarded in the (1.6 1.6 0) once having subtracted the analyser crystal leakage (not shown here). The left panel shows that there exist both a resonant and off- resonant signal for $\mathbf{q}_{w,2-}$ in the whole studied range of φ. The intensity at the resonance is ~ $10^{-6}$ that of the Bragg (220) reflection at the same energy. By analogy to checkerboard model as established for the half- doped compound, this reflection would be the counterpart of the (h/2 h/2 0), with h odd, associated to CO. The results indicate the existence of a small charge disproportionation between Mn sites and a Thomson contribution due to little shifts of atoms from their ideal crystallographic positions in the *I4/mmm* system. The intensity at the maximum of the main resonance (after off- resonant contribution subtraction) shows little variations with the azimuthal angle as plotted in the inset. This is in contrast to the observations reported in other half- doped compounds [10]. These changes could be ascribed to either the different dimensionality or the incommensurate character of this reflection. In the right panel, $\mathbf{q}_{v,1(2)\pm}$~(0.8 0.8 0), (1.2 1.2 0), (1.8 1.8 0) and (2.2 2.2 0) are shown after normalisation to the same scale. Following the analogy to the x=0.5 case, these would be ATS reflections [20] whose origin might be in the $x^2-z^2/y^2-z^2$ OO. The identical main resonances confirm their same dipolar character while at the pre-peak a **Q** dependence is observed. Thus, it is interesting to note that its shape can be grouped into couples, $q_{v,1+} \sim q_{v,2+}$ and $q_{v,1-} \sim q_{v,2-}$. On the other hand, the inset shows that the azimuthal evolution at both the pre-peak and main peak follows a $\sim \sin^2 \phi$ dependence with coincident zero minima as reported for the half- doped compound [3].

Figure 2 plots temperature dependences of the diffraction angle $2\theta$ of the superlattice (1.8 1.8 0) and the Bragg (2 2 0) atomic reflections. First, OO-kind reflection (1.8 1.8 0) peak is broadened, which is indicative of a shorter correlation length than for nuclear reflections. Second, we notice on their commensurability. The contraction of *a* tetragonal lattice parameter when lowering T is reflected in the drift of the angular position of (220) reflection to larger angles, about 0.2° from 240 to 100 K. In contrast, any variation is hardly detectable in (1.8 1.8 0). This would mean that Miller indexes are incommensurate to the crystal lattice and T dependent. Indeed, the actual (h k l) found in the experiment for $q_v$ and $q_w$ are typically deviated $\sim 10^{-2}$ Å$^{-1}$ from ideal rational values. A lock- in to the latter has not been observed in the range from 80 to 220 K. It turns out evident that these results make difficult the direct applicability of the checkerboard model as understood for manganites with x=0.5 but rather agree with a picture where the electronic charge density is strongly coupled to the lattice, propagating as a CDW.

We do not discuss here on the magnitude of the formal charge segregation between inequivalent Mn sites. It remains as the subject for a future deeper analysis.

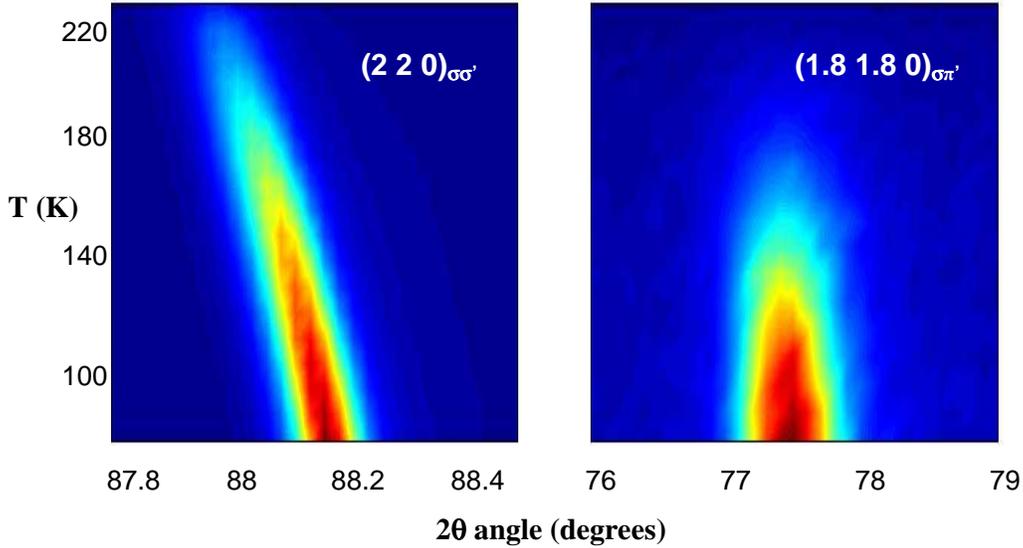

Figure 2. Temperature dependence of $2\theta$ angular position in (220) Bragg and (1.8 1.8 1.8) superlattice reflections. Color (arbitrary units) varies linearly with intensity from blue to red scale.

In the case of La(Pr)$_{1/3}$Sr$_{2/3}$FeO$_3$, the variation of $2\theta$ in $(4/3\ 4/3\ 4/3)_p$ superstructure confirms the commensurate character of the CO phase (not shown here). In Figure 3 (a) we see the $q'_{CO,H}=(h'/3\ h'/3\ h'/3)_p$ energy scans in the $\sigma$-$\sigma'$ channel for h'=2, 4, 5 in PSFO. Resonances are observed at the Fe K edge energy for h'=4, 5, while $(2/3\ 2/3\ 2/3)_p$ shows a "valley". Off- resonant intensities are also highly dependent on **Q**. It is maximal for h'= 4 and much weaker for h'= 2, 5. The comparison to LSFO [21] reveals a similar shape in all spectra though resonances (h'=4, 5) are less marked in Pr sample and relative intensity for h'=2 is quite different. A smaller resonant signal correlates to previous results arguing that the charge segregation between different Fe sites decreases proportionally to $R$ cation size [22]. We recently demonstrated that a lattice modulation involving not only Fe but also La(Sr) and O atoms is necessary in order to reproduce RXS spectra in LSFO. Furthermore, these results combined to x- ray absorption data permitted to discard the integer CO model based on a …+3+3+5+3+3+5… sequence. Simulations converged to a real average charge disproportionation of 0.6 e⁻, i.e. formal $Fe^{+3.3}$ and $Fe^{+3.9}$ ordered along the [111] cubic direction in a 1:2 ratio in combination with a displacement of La(Sr)O atomic planes in antiphase to Fe [21,23]. Spectra in figure 3 suggest that a similar but not identical structural model would be necessary to fit PSFO curves. All studied superlattice reflections disappear at T~180 K.

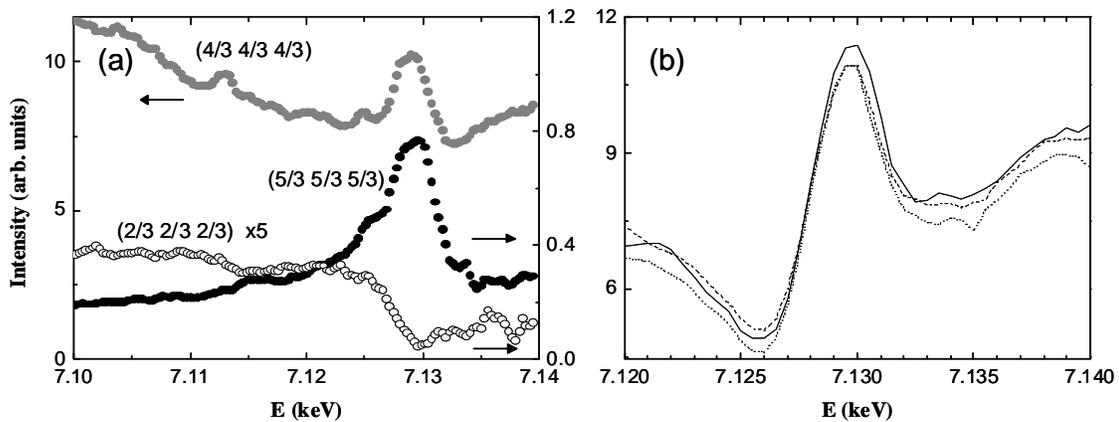

Figure 3. (a) Energy dependence of superlattice reflections in PSFO at 100 K. (b) Energy profile of LSFO (4/3 4/3 4/3)$_p$ for $\phi$=-15 (solid), 30 (dashed) and 75 (dotted) degrees.

Compared to Mn oxides, no OO- kind superstructure has been found in $R_{1/3}Sr_{2/3}FeO_3$ ferrites. Moreover, energy scans of the (h'/3 h'/3 h'/3)$_p$ reflections are identical as a function of the azimuth angle in $R_{1/3}Sr_{2/3}FeO_3$ ferrites, as demonstrated in figure 3(b). In manganites, the appearance of the OO- kind superstructure is usually explained in terms of the partial occupation of lowest energy $e_g$ orbitals, in direct correlation to the development of a Jahn- Teller distortion in Mn atoms that increases a Mn-O pair bond length. This type of distortion seems to be absent in ferrites. The description of $Fe^{4+}$ ($3d^4$) as a low spin system without $e_g$ electrons is in agreement. The comparison of CO/OO phases in layered and non- layered samples can help to decouple the role of geometrical and electronic order parameters. Actually, due to the different dimensionality of the lattice, the $e_g$ orbitals involved in the OO formation in $R_{0.5}A_{0.5}MnO_3$ and $La_{0.5}Sr_{1.5}MnO_4$ are thought to be different [24].

### 4. Conclusions

We have investigated the charge and orbital ordered phases in $La_{0.4}Sr_{1.6}MnO_4$ and $La(Pr)_{1/3}Sr_{2/3}FeO_3$ by RXS at the Mn and Fe K absorption edges. The transition metal is in a formal intermediate oxidation state in both samples, i.e $Mn^{+3.60}$ and $Fe^{+3.67}$, respectively. In the case of the Mn oxide, recorded CO and OO spectra show reminiscences from the well known half- doped (x=0.5) parent compound. However, in comparison to the latter, peaks intensities are strongly reduced (a factor ~$10^2$) and considerably broadened (unpublished data) due to a shorter correlation length. A Mn bimodal distribution model considering charge- stacking faults or COO phase correlated "melting" islands would be then plausible. Though, the incommensurate character these superlattice reflections appear to have (fig.2) seems hardly reconcilable with a model where CO is only supported by Mn atoms. Other experimental results point also to discard this possibility and support a continuous charge modulation, i.e. a CDW model [14]. This could commensurate to the lattice for particular rational values of hole doping, as $x=1/2$ and $2/3$.

CO is also observed in $R_{1-x}S_xFO$ in a wide range of hole- doping, where similar commensurate-incommensurate coupled electron- lattice modulations have been proposed to explain the observed phenomenology. For $x=2/3$, we have shown that Pr based sample modulated crystal structure in the CO phase must be very similar to that we proposed for the La sample [21]. Here we argued that small cooperative displacements of Fe, La(Sr) and O atoms are necessary to account for the measured RXS spectra and that Fe charge segregation was not very significant.

We realise that the rich phenomenology of COO phases in mixed valence transition metal oxides still leaves open questions, and further experimental and theoretical efforts must be done in order to improve the actual picture.

**Acknowledgements**
The authors wish to thank ESRF for beamtime granting and financial support from Spanish MICINN FIS2008-03951 and DGA Camrads projects.